\pdfoutput=1
\documentclass[fleqn,usenatbib,useAMS,twocolumn]{mnras}

\usepackage{graphicx}	
\usepackage{amsmath}	
\usepackage{amssymb}	
\usepackage{multicol}        
\usepackage{bm}		
\usepackage{pdflscape}	
\usepackage{epstopdf}
\usepackage[T1]{fontenc}
\usepackage{ae,aecompl}

\newcommand{\ms}[1]{\textcolor{black}{#1}}

\newcommand{\beq}{\begin{equation}}
\newcommand{\beqa}{\begin{eqnarray}}
\newcommand{\eeq}{\end{equation}}
\newcommand{\eeqa}{\end{eqnarray}}

\newcommand{\simgt}{\lower.5ex\hbox{$\; \buildrel > \over \sim \;$}}
\newcommand{\simlt}{\lower.5ex\hbox{$\; \buildrel < \over \sim \;$}}

\newcommand{\bd}[1]{\mbox{\boldmath $#1$}}

\usepackage[T1]{fontenc}
\usepackage{ae,aecompl}

\usepackage{newtxtext,newtxmath}

\title[Radio sources and CIB in tSZ-WL correlation]{Impact of radio sources and cosmic infrared background on 
thermal Sunyaev-Zel'dovich - gravitational lensing cross correlation}

\author[M. Shirasaki]{
Masato Shirasaki$^{1}$\thanks{E-mail: masato.shirasaki@nao.ac.jp}
\\
$^{1}$National Astronomical Observatory of Japan, Mitaka, Tokyo 181-8588, Japan
}

\date{Accepted XXX. Received YYY; in original form ZZZ}

\pubyear{2018}

\begin{document}
\label{firstpage}
\pagerange{\pageref{firstpage}--\pageref{lastpage}}
\maketitle

\begin{abstract}
Cross correlation with thermal Sunyaev-Zel'dovich (tSZ) effect in cosmic microwave background observation and weak gravitational lensing effect in galaxy imaging survey opens 
a new window on constraining matter contents in the Universe at redshifts less than 1.
In this paper, we study the impact of radio sources and cosmic infrared background (CIB) on observed tSZ-lensing correlation.
Assuming the best-fit model of CIB by the {\it Planck} satellite, we estimate that 
the residual correlation of CIB with large-scale structures will be 
of an order of $2\%$ of expected tSZ-lensing correlation from intracluster medium in current lensing surveys.
On the other hand, \ms{despite large model uncertainties,} we find that correlation of lensing and radio sources can induce a {\it negative} correction for the observed tSZ-lensing correlation with a $\sim10-30\%$ level.
This is originated from positive cross correlation with radio sources and lensing at $\sim100\, {\rm GHz}$ frequency, 
whereas tSZ-lensing correlation should show a negative value in temperature fluctuations at that frequency. 
We also show that the negative correction by radio-lensing correlation 
can solve the tension between recent measurements of tSZ-lensing correlation and an expected signal from ``universal'' gas pressure profile of nearby galaxy clusters,
when the radio sources with flat-spectral index are assumed to 
populate massive cluster-sized dark matte halos.
Our results indicate that minor source population in radio bands can play an important role in determining observed tSZ-lensing correlation at $\simlt10$ arcmin.
\end{abstract}




\section{Introduction}

The observation of cosmic microwave background (CMB) radiation by 
the {\it Planck} satellite\footnote{\url{https://www.cosmos.esa.int/web/planck}} 
is among the most crucial for modern astrophysics and cosmology.
A wide frequency coverage in the {\it Planck} observation enables to not only have a robust estimate of CMB \citep[e.g.][]{2016A&A...594A...9P}, but also study other interesting physical effects printed in CMB radiation.
Inverse Compton scattering of CMB photons by hot relativistic electrons \citep{1969Ap&SS...4..301Z}, referred to as thermal Sunyaev-Zel'dovich (tSZ) effect,
is one of the main science targets in the {\it Planck} mission \citep[e.g.][]{2016A&A...594A..22P}.
Since the tSZ effect leads to a frequency-dependent distortion of  the CMB black-body spectrum \citep{1969Ap&SS...4..301Z},
the measurement on different wavelengths is essential as the {\it Planck} satellite has done.
Previous studies have shown that 
tSZ effect by intracluster medium (ICM) in galaxy clusters 
is of extreme cosmological importance.
The statistics of tSZ effect can probe the abundance of most massive dark matter halos at various redshifts and 
show a strong dependence on the amplitude in linear density fluctuations \citep[e.g.][]{1999ApJ...526L...1K,
2002MNRAS.336.1256K, 2012ApJ...760....5B}.
It is worth noting that the precise measurement of the amplitude in density fluctuations allows us to constrain physics beyond the standard model, 
including massive neutrinos 
\citep[e.g.][]{2008PhRvL.100s1301S}, 
dark matter 
\citep[e.g.][]{2011PhRvD..84f3507S, 2015JCAP...09..067E}, 
and cosmic acceleration
\citep[e.g.][]{2013PhR...530...87W}

To use tSZ effect for cosmological analyses, 
we need accurate modeling of ICM.
The tSZ statistics have degeneracy between 
the ICM property and cosmological parameters in principle 
\citep[e.g.][]{2010ApJ...725...91B}.
Hence, the current cosmological constraints by the tSZ statistics 
hinge on the still poorly understood property of ICM
\citep[e.g][]{2016A&A...594A..22P, 2017MNRAS.469..394H}.
Various approaches have been proposed to learn 
the ICM physics more and 
break the degeneracy in tSZ statistics with astronomical measurements other than tSZ effect
\citep[e.g.][]{2017JCAP...11..040B,2018PhRvD..97h3501H}.
Among them, the measurement of gravitational lensing effect in galaxy imaging survey can play a central role in determining the relation between the ICM and large-scale structures.

Gravitational lensing effect causes a small distortion 
in shape of distant sources and its amplitude 
depends on surface mass density in the direction of individual sources \citep[e.g][for review]{Bartelmann2001}. 
Therefore, the information of total matter density 
(including dark matter) can be obtained with 
the lensing measurement, and those information 
would be a base for developing accurate model of ICM.
\citet{2015ApJ...812..154B} have shown that
the cross correlation between tSZ and lensing can probe 
the ICM over a wide range of halo masses and redshifts
and future measurement of the cross correlation
can constrain the relation between ICM and dark matter halos
with a 5-20\% precision.

Recently, the measurements of tSZ-lensing cross correlation
have been performed with lensing data in the Canada France Hawaii Lensing Survey\footnote{\url{http://www.cfhtlens.org/}} \citep[CFHTLenS:][]{2014PhRvD..89b3508V} and 
the Red Cluster Sequence Lensing Survey\footnote{\url{http://www.rcslens.org/}} \citep[RCSLenS:][]{2017MNRAS.471.1565H}.
The signal-to-noise ratio in the latest measurement already reaches a $13-17\sigma$ level \citep{2017MNRAS.471.1565H}, implying
that ongoing and future lensing surveys will be able to present more precise measurements.
A natural question then arises: can we explain the observed tSZ-lensing correlation in CFHTLenS and RCSLenS assuming the ICM properties observed in nearby clusters \citep[e.g.][]{2013A&A...550A.131P}? This question is still under debate.
At least, it seems difficult to explain the \ms{tSZ-lensing} correlation at angular scale of $<10$ arcmin by the model with the ICM property from local observation \citep{2015JCAP...09..046M}. 
To fill the gap, one may need a relatively small amplitude 
in matter density fluctuations today compared to the constraint by CMB measurements and/or a strong baryonic feedback by active galactic nuclei (AGN) 
in galaxy groups \citep{2017MNRAS.471.1565H}.
Furthermore, \citet{2018MNRAS.475..532O} examined a semi-analytic model of ICM in \citet{2010ApJ...725.1452S} 
to explain both of the tSZ-lensing correlation in \citet{2017MNRAS.471.1565H} and the power spectrum of tSZ effect measured in \citet{2016A&A...594A..22P}. The authors found that the tSZ power spectrum prefers $\sim30\%$ of ICM pressure at outer region in galaxy clusters to be non-thermal, while the tSZ-lensing correlation prefers a much lower fraction of non-thermal pressure with a level of $\simlt5\%$.
At present, the ICM model in \citet{2010ApJ...725.1452S} cannot explain the both statistics simultaneously without reducing gas fraction in group-scaled halos.

Motivated by the inconsistency between the observed tSZ-lensing
signal and the model with the ICM property of nearby clusters constrained by {\it Planck} \citep{2013A&A...550A.131P},
we here consider other relevant effects in the cross correlation measurement.
In this paper, we study the impact of the presence of astrophysical sources in brightness temperature maps on the cross correlation measurement. 
Except for galactic sources, relevant sources in the cross correlation measurement are expected to be radio point sources
and cumulative emission from star-forming dusty galaxies, referred to as cosmic infrared backgrounds (CIB).
The impact of those extragalactic sources 
on construction of tSZ map has been studied in \citet{2014A&A...571A..21P} and \citet{2016A&A...594A..22P},
while it still remains uncertain if one can ignore the radio sources and CIB in the tSZ-lensing cross correlation
\citep[see][for the study with a similar motivation]{2015A&A...575L..11H}.
We work with a simple halo-model approach to predict 
possible contamination in the tSZ-lensing correlation induced
by relevant astrophysical sources living in large-scale structures at various redshifts.
Using our model, we evaluate the corrections for the observed tSZ-lensing correlation signal by CIB- and
radio-lensing correlations.

This paper is organized as follows.
In Section~\ref{sec:tsz_estimation}, we first describe the methodology for construction of tSZ effect
from multiple-frequency CMB measurements and discuss the possible effects from astrophysical sources at frequency bands of interest. 
We then summarize the basics of gravitational lensing
and the cross correlation between tSZ and lensing in Section~\ref{sec:cross_tSZ_lens}.
Our model of the tSZ-lensing correlation is summarized 
in Section~\ref{sec:model} and Section~\ref{sec:result} presents the outcome of our model and comparison of 
the observed tSZ-lensing correlation with the model.
We conclude the paper in Section~\ref{sec:conc}.
Throughout this study, we adopt the flat-geometry $\Lambda$CDM (Lambda cold dark matter) model that is consistent with the {\it Planck} 2015 results \citep[see the result of ``TT+lowP+lensing'' in ][]{2016A&A...594A..13P}. 
The cosmological parameters are as follows: the matter density, $\Omega_{\rm m0}=1-\Omega_{\Lambda}=0.308$, the baryon density, $\Omega_{\rm b0}=0.0484$, the Hubble parameter, $h=H_0/(100 {\rm km} \, {\rm s}^{-1} \, {\rm Mpc}^{-1}) = 0.678$, the present amplitude of density contrast at $8 \, h^{-1} \, {\rm Mpc}$, $\sigma_8=0.8149$, and the spectral index, $n_{\rm s}=0.9677$.  

\section{Estimation of tSZ effect from CMB measurements}
\label{sec:tsz_estimation}

At frequency $\nu$, the change in CMB temperature by the tSZ effect is expressed as
\beq
\frac{\Delta T}{T_0} = g(x) y, \label{eq:T-y}
\eeq
where $T_{0}=2.725\, {\rm K}$ is the CMB temperature \citep{2009ApJ...707..916F}, 
$g(x) = x{\rm coth}(x/2)-4$ with $x=h\nu/k_{\rm B}T_{0}$,
$h$ and $k_{\rm B}$ are the Planck constant and the Boltzmann constant, respectively\footnote{
In this paper, we ignore the relativistic correction for $g(x)$ which is only important 
for the tSZ effects in most massive galaxy clusters \citep{1998ApJ...502....7I, 1998ApJ...508...17N}.
Note that the tSZ-lensing correlation can probe the ICM at clusters with masses of 
$\sim10^{14}\, h^{-1}M_{\odot}$ \citep{2015JCAP...09..046M, 2018MNRAS.475..532O}.
}.
Compton parameter $y$ is computed as the integral of the electron pressure $P_{\rm e}$ along a line of sight:
\beq
y(\bd{\theta}) = \int_0^{\chi_H}\, \frac{{\rm d}\chi}{1+z}\, \frac{k_{\rm B}\sigma_{\rm T}}{m_{\rm e}c^2} P_{\rm e}\left(r(\chi)\bd{\theta}, z(\chi)\right), \label{eq:tSZ_y}
\eeq
where $\sigma_{\rm T}$ is the Thomson cross section, $\chi$ is the comoving radial distance to redshift $z$, 
$r(\chi)$ is the angular diameter distance,
and $\chi_{H}$ is the comoving distance up to $z\rightarrow \infty$.

Given the frequency dependence in the tSZ effect as Eq.~(\ref{eq:T-y}), 
one can construct an estimator of Compton $y$ map from brightness temperature maps at multiple frequencies as
\beq
\hat{y}(\bd{\theta}) = \sum_{i} w_{i} \frac{T_{i}(\bd{\theta})}{T_{0}}, \label{eq:ILC}
\eeq
where $\hat{y}$ is an estimated Compton $y$ parameter, $T_{i}$ is the observed temperature at $i$-th frequency 
$\nu_{i}$ and the sum in Eq.~(\ref{eq:ILC}) is over frequency bands of interest.
\citet{2014PhRvD..89b3508V} imposed three conditions to construct the Compton $y$ map for the cross correlation with the tSZ effect and 
weak lensing effect in galaxy shapes. Those include (i) $\sum_{i} g(x_{i}) w_{i} = 1$ to produce an unbiased Compton $y$ map on average,
(ii) $\sum_{i} w_{i} = 0$ to null the primary CMB fluctuations, and (iii) $\sum_{i} w_{i} \cdot c_{i} \nu_{i}^{\beta_{d}} = 0$
to remove a contribution from dust emission with spectral index $\beta_{d}\sim2$ in antenna temperature units
(The factor $c_i$ is given by the conversion of antenna temperature to thermodynamic temperature. See also Section~\ref{subsec:astro}).
In \citet{2014PhRvD..89b3508V}, the authors worked on the {\it Planck} temperature maps \citep{2014A&A...571A...6P} at four frequency bands
of 100, 143, 217, and 353 GHz to determine the weight $w_{i}$ for different $\beta_{d}$.
Throughout this paper, we work with the weight in the $y$ map named {\it Planck} C in \citet{2014PhRvD..89b3508V} for representative example.

In addition to estimating the tSZ effect, \citet{2014PhRvD..89b3508V} examined to set different weights in Eq.~(\ref{eq:ILC})
by projecting out the both of tSZ effect and primary CMB fluctuation (i.e. $\sum_{i} g(x_{i}) w_{i} = \sum_{i} w_{i} = 0$).
Even after nulling the tSZ effect in Eq.~(\ref{eq:ILC}), they found a weak correlation of the observed $\hat{y}$ and weak lensing (see Figure 4 in \citet{2014PhRvD..89b3508V}). This supports that 
the contamination in $y$ map construction should (partly) arise from some extragalactic sources since weak gravitational lensing effect
is relevant for the large-scale structures at $z\sim0.2-0.4$ \citep[][also see Figure~\ref{fig:dlnIdz_Wkappa}]{2013MNRAS.433.3373V}. 
In this paper, we study the correlation of $\hat{y}$ and gravitational lensing 
when setting weight in Eq.~(\ref{eq:ILC}) so as to remove the tSZ effect\footnote{
\ms{The conditions in \citet{2014PhRvD..89b3508V} for nulling tSZ cannot determine the overall normalization of weights. This degree of freedom does not affect our results since we use the same weights as in \citet{2014PhRvD..89b3508V}.}
}. 
For the weight so as to null the tSZ effect in $\hat{y}$ map, we work with the case named as {\it Planck} E' in \citet{2014PhRvD..89b3508V}.

Note that the $\it Planck$ team has worked on more sophisticated approach than Eq.~(\ref{eq:ILC}) to construct a Compton $y$ map. They allowed a spatially varying weight in Eq.~(\ref{eq:ILC}) \citep[see][for the latest map]{2016A&A...594A..22P}.
For a spatially varying weight, we can easily take into account the effect on the cross correlation measurements 
by using the formula in the Appendix A of \citet{2014A&A...571A..21P}. The primary purpose in this paper is to follow the analyses in \citet{2014PhRvD..89b3508V}, and we leave it to future work to include the effect of spatially varying weight.

Nevertheless, the following results are expected to be less affected by the choice of the weight in Eq.~(\ref{eq:ILC}) or spatially varying weight in \citet{2016A&A...594A..22P}. First of all, the spatially varying weights can reduce the contamination from spatially-resolved objects \citep{2013A&A...558A.118H}, but it is no longer valid for unresolved sources such as CIB or faint radio sources. Also, the spatially varying weight in \citet{2016A&A...594A..22P} is introduced so as to localize tSZ effects in temperature maps at different frequency channels.  
Hence, the frequency dependence of such weight is expected to be similar to one of the weight in Eq.~(\ref{eq:ILC}).
In addition, the weight in \citet{2016A&A...594A..22P} uses small-scale information of temperature maps to localize the tSZ effects,
leaving the impact of faint localized objects on the cross correlation analysis.

\section{Cross correlation with Compton y and weak lensing}
\label{sec:cross_tSZ_lens}

As introduced in Section~\ref{sec:tsz_estimation}, the actual observable $\hat{y}$ is given by the linear combination of brightness temperature as in Eq.~(\ref{eq:ILC}). 
Here we summarize the basic of the cross correlation between $\hat{y}$ map and 
weak gravitational lensing effect in galaxy imaging survey. 
Lensing convergence $\kappa$ is responsible for the strength of weak gravitational lensing effect.
Under the Born approximation, 
one can express the lensing convergence field as the weighted 
integral of matter overdensity field $\delta_{\rm m}(\bd{x})$ \citep{Bartelmann2001}:
\beq
\kappa(\bd{\theta})
= \int_{0}^{\chi_{H}} {\rm d}\chi \ W_{\kappa}(\chi)\delta_{\rm m}(r(\chi)\bd{\theta}, z(\chi)), \label{eq:kappa_delta}
\eeq
where $W_{\kappa}(\chi)$ is called lensing kernel.
For a given redshift distribution of source galaxies,
the lensing kernel is expressed as
\beq
W_{\kappa}(\chi) = \frac{3}{2}\left( \frac{H_{0}}{c}\right)^2 \Omega_{\rm m0} \frac{r(\chi)}{a(\chi)}\, \int_{\chi}^{\chi_{H}} {\rm d}\chi^{\prime} p(\chi^{\prime})\frac{r(\chi^{\prime}-\chi)}{r(\chi^{\prime})}, \label{eq:lens_kernel}
\eeq
where $p(\chi)$ represents the redshift distribution of source galaxies
normalized to $\int {\rm d}\chi \, p(\chi) =1$.
Hence, the quantity of interest in this paper is defined as
\beq
\xi_{y-\kappa}(\theta) 
\equiv \langle \hat{y}(\bd{\phi}) \kappa(\bd{\phi}+\bd{\theta})\rangle
= \sum_{i} w_{i}/T_0 \langle T_{i}(\bd{\phi}) \kappa(\bd{\phi}+\bd{\theta})\rangle,
\eeq
and the cross correlation of $\xi_{y-\kappa}$ can be computed as \citep[e.g.][]{2017MNRAS.471.1565H}
\beq
\xi_{y-\kappa}(\theta) = \sum_{\ell} \left(\frac{2\ell+1}{4\pi}\right) C_{y-\kappa}(\ell) P_{\ell}(\cos\theta) b^{\kappa}_{\ell} b^{y}_{\ell},
\eeq
where $P_{\ell}$ are the Legendre polynomials,
$b^{\kappa}_{\ell}$ and $b^{y}_{\ell}$ are the smoothing kernel of $\kappa$ and $\hat{y}$ maps, respectively.
The power spectrum of $C_{y-\kappa}(\ell)$ can be decomposed into
\beq
C_{y-\kappa}(\ell) = \left(\sum_{i} w_{i} g(x_i) \right) C_{y-\kappa}({\rm tSZ}|\ell) + \sum_{i} w_{i} C_{\kappa-T}(\nu_{i}|\ell), \label{eq:obs_yk_power}
\eeq
where the first term in the right hand side expresses the correlation of tSZ effect and lensing convergence as studied in the literature,
whereas the second term is new contribution arising from the correlation of astrophysical sources in the {\it Planck} bands
and lensing convergence. Note that the second term in the r.h.s of Eq.~(\ref{eq:obs_yk_power}) can have 
different dependence in frequency $\nu$ from tSZ effect and CMB black-body spectrum, and it cannot be vanished in general.
We define $C_{\kappa-T}(\nu_{i}|\ell)$ to be dimensionless by normalizing the $T$ field with CMB temperature $T_{0}$ throughout this paper.

\section{Model}
\label{sec:model}

In this section, we describe a theoretical model of Eq.~(\ref{eq:obs_yk_power}) based on halo-model approach.

\subsection{Intracluster medium}
\label{subsec:ICM}

We first summarize the halo-model prediction of tSZ-lensing cross correlation induced by ICM as developed in \citet{2014JCAP...02..030H, 2015JCAP...09..046M}. In the halo model, we can decompose the power spectrum into two components as:
\beq
C_{y-\kappa} = C^{\rm 1h}_{y-\kappa} + C^{\rm 2h}_{y-\kappa},
\eeq
where the first term in the r.h.s arises from 
the correlation within single halos, while the second term represents the correlation due to clustering of neighboring halos.
For tSZ effect, one can express those terms as
\begin{align}
C^{\rm 1h}_{y-\kappa}({\rm tSZ}|\ell)
=& \int_{0}^{z_{\rm max}}\, {\rm d}z\, \frac{{\rm d}V}{{\rm d}z
{\rm d}\Omega}
\int_{M_{\rm min}}^{M_{\rm max}}\, {\rm d}M\, \frac{{\rm d}n}{{\rm d}M}\, y_{\ell}(M,z)\, \kappa_{\ell}(M,z), \label{eq:tsz_WL_1h} \\
C^{\rm 2h}_{y-\kappa}({\rm tSZ}|\ell)
=& \int_{0}^{z_{\rm max}}\, {\rm d}z\, \frac{{\rm d}V}{{\rm d}z
{\rm d}\Omega}\, P_{\rm L}(k=\ell/\chi,z) \nonumber \\
&
\,\,\,\,\,\,
\,\,\,\,\,\,
\,\,\,\,\,\,
\times
\left[
\int_{M_{\rm min}}^{M_{\rm max}}\, {\rm d}M\, \frac{{\rm d}n}{{\rm d}M}\, y_{\ell}(M,z)\, b(M,z)
\right]
\nonumber \\
&
\,\,\,\,\,\,
\,\,\,\,\,\,
\,\,\,\,\,\,
\times
\left[
\int_{M_{\rm min}}^{M_{\rm max}}\, {\rm d}M\, \frac{{\rm d}n}{{\rm d}M}\, \kappa_{\ell}(M,z)\, b(M,z)
\right], \label{eq:tsz_WL_2h}
\end{align}
where 
we set $z_{\rm max}=7$, $M_{\rm min}=10^{10}\, h^{-1}M_{\odot}$ and $M_{\rm max}=10^{16}\, h^{-1}M_{\odot}$,
$P_{\rm L}(k,z)$ is the linear matter power spectrum,
${\rm d}n/{\rm d}M$ is the halo mass function, and $b$ is the linear halo bias. In this paper, we define the halo mass $M$ 
by spherical overdensity (SO) with respect to 200 times mean matter density. 
We adopt the model of halo mass function in \citet{2008ApJ...688..709T} and linear bias in \citet{2010ApJ...724..878T}.
In Eqs.~(\ref{eq:tsz_WL_1h}) and (\ref{eq:tsz_WL_2h}),
$\kappa_{\ell}$ is the Fourier transform of lensing convergence profile of single dark matter halo with the NFW density profile \citep{1996ApJ...462..563N}:
\beq
\kappa_{\ell}(M, z) = 
\frac{W_{\kappa}(\chi(z))}{\chi^2} \int\, {\rm d}r\, 4\pi r^2\, \frac{\sin(\ell r/\chi)}{\ell r/\chi} \frac{\rho_{\rm NFW}(r, M, z)}{\bar{\rho}_{\rm m}},
\eeq
where $\rho_{\rm NFW}$ is the NFW profile 
and $\bar{\rho}_{\rm m}$ is the mean matter density in the Universe. The NFW profile can be characterized by single parameter called halo concentration for a given SO mass. In this paper, we adopt the model of halo concentration developed in \citet{2015ApJ...799..108D}.

Similarly, we define $y_{\ell}$ in Eqs.~(\ref{eq:tsz_WL_1h}) and (\ref{eq:tsz_WL_2h}) as the Fourier transform of Compton $y$ profile in single halo (see Eq.~\ref{eq:tSZ_y}):
\beq
y_{\ell}(M,z) = 
\frac{4\pi r_{\rm 500}}{\ell^2_{\rm 500}}
\frac{\sigma_{\rm T}}{m_{\rm e}c^2}
\int\, {\rm d}x\, x^2\, \frac{\sin(\ell x/\ell_{500})}{\ell x/\ell_{500}} P_{{\rm e},h}(x, M, z),
\eeq
where $r_{500}$ is the SO radius 
with respect to 500 times critical density, 
we define as $x = ar/r_{500}$ and $\ell_{500}=a\chi/r_{500}$.
When computing $y_{\ell}$, we adopt the model of  
3D electron pressure profile in single halo $P_{{\rm e},h}$
as constrained in \citet{2013A&A...550A.131P},
\begin{align}
P_{{\rm e}, h}(x=r/r_{500}, M, z) =& 
1.65 \times 10^{-3}\,  \left[{\rm keV}\, {\rm cm}^{-3}\right]\, E^{8/3}(z) \nonumber \\
&
\times
\left(\frac{M_{500}}{3\times10^{14}\, h_{70}^{-1}M_{\odot}}\right)^{2/3+0.12}\, P(x)\, h_{70}^2, \label{eq:UPP_PLANCK}
\end{align}
where $E(z)=H(z)/H_{0}$, $h_{70}=H_0/70$, $M_{500}=4/3\pi\,\cdot 500\,\rho_{\rm crit}\,r^3_{500}$ ($\rho_{\rm crit}$ is the critical density in the Universe) and $P(x)$ is so-called universal pressure profile 
\citep{2007ApJ...668....1N}. The functional form of $P(x)$ is given by
\beq
P(x) = \frac{P_0}{(c_{500}x)^{\gamma}\left[1+(c_{500}x)^{\alpha}\right]^{(\beta-\gamma)/\alpha}},
\eeq
where we adopt the best-fit values of five parameters 
($P_0, c_{500}, \alpha, \beta$, and $\gamma$) from \citet{2013A&A...550A.131P}.
Note that the input mass parameter $M_{500}$ in Eq.~(\ref{eq:UPP_PLANCK}) will be affected by hydrostatic mass bias.
For a given halo mass of $M$ (the SO mass w.r.t $200$ times mean matter density), we compute $M_{500}$ by using the halo concentration as in \citet{2003ApJ...584..702H} and then include the hydrostatic mass bias $b_{\rm HM}$ by 
$M_{500} \rightarrow (1-b_{\rm HM})M_{500}$ and $r_{500} \rightarrow (1-b_{\rm HM})^{1/3} r_{500}$
for Eq.~(\ref{eq:UPP_PLANCK}). 
We set $1-b_{\rm HM}=0.833$ as follows in \citet{2016MNRAS.463.1797D}.

It is worth noting that \citet{2016MNRAS.463.1797D} have shown the above ICM model can explain the observed tSZ power spectrum \citep{2016A&A...594A..22P}. In fact, one can also explain the tSZ-lensing correlations \citep{2014PhRvD..89b3508V, 2017MNRAS.471.1565H} with the above ICM model by setting $P_{0}\sim4.4$ (the best-fit value from {\it Planck} analysis is 6.41), whereas it turns to be difficult to explain the tSZ power spectrum. In addition, the value of $P_{0}\sim4.4$ is inconsistent with the observations of nearby galaxy clusters \citep{2010A&A...517A..92A, 2013A&A...550A.131P}.

\subsection{Astrophysical sources}
\label{subsec:astro}

Next we consider cumulative emission from astrophysical sources at frequency of $100-800$ GHz
and the cross correlation with lensing convergence field.
At the frequency of interest, relevant astrophysical sources include point sources in radio bands,
referred to as radio galaxies or/and radio AGN in the literature, and CIB emission.

Observed specific intensity at a given frequency $\nu$ can be expressed as
\beq
I_{\nu} (\bd{\theta}) = \int\, \frac{{\rm d}\chi}{1+z} j_{\nu}(r(\chi)\bd{\theta}, z(\chi)),
\eeq
where $j_{\nu}(\bd x)$ represents the comoving specific emission coefficient.
One can convert the specific intensity to antenna temperature using the CMB black-body spectrum as
\begin{align}
T_{\nu}(\bd{\theta}) =& \left(\frac{\partial B_{\nu}}{\partial T}\Bigg|_{T=T_0}\right)^{-1} I_{\nu}(\bd{\theta}), \\
\frac{\partial B_{\nu}}{\partial T}\Bigg|_{T=T_0}
=& 99.27 \left[{\rm Jy}\, {\rm str}^{-1} / \mu{\rm K}\right]\,  \frac{x^4\, e^x}{(e^x-1)^2}, 
\end{align}
where $x=h\nu/k_{\rm B}T_{0} = \nu/56.84\, \rm GHz$.

In the halo-model approach, we can relate the emission coefficient with underlying astrophysical sources as follows \citep[e.g.][]{2012MNRAS.421.2832S}:
\beq
j_{\nu} = \int {\rm d}L\, \frac{{\rm d}n}{{\rm d}L}(L,z)\, \frac{L_{(1+z)\nu}}{4\pi}, \label{eq:emission_coef}
\eeq
where $L$ is the luminosity in infrared or radio for our case, 
${\rm d}n/{\rm d}L$ represents the luminosity function, and 
$L_{(1+z)\nu}$ is related to the flux $S_{\nu}$ from relevant object as
\beq
S_{\nu} = \frac{(1+z)^{-1}\, L_{(1+z)\nu}}{4\pi\,\chi^2}.
\eeq
\ms{In the Appendix, we summarize the derivation for cross power spectra of weak lensing and astrophysical sources based on the halo model.}

\subsubsection*{Radio sources}

For radio sources, we work with three-population model as introduced in \citet{2005A&A...431..893D}.
In this model, extragalctic radio sources consist of flat-spectrum radio quasars (FSRQs), BL Lac objects, 
and steep-spectrum sources. Their spectral index is assumed to be $\alpha=0.1$, $0.1$, and $0.7$ for FSRQs, BL Lac objects, and steep-spectrum sources, respectively (the index is defined in terms of $S_{\nu}\propto \nu^{-\alpha}$).
The radio luminosity function at 1.4 GHz for these three populations has been constrained in \citet{2010MNRAS.404..532M}
with local luminosity functions, multifrequency source counts and redshift distributions.
In this paper, we adopt the radio luminosity function in \citet{2010MNRAS.404..532M} and 
assume that the emission coefficient from extragalactic radio sources can be computed as
\begin{align}
j_{\nu} \simeq& \sum_{i} \int {\rm d}L_{1.4}\, \frac{{\rm d}n_i}{{\rm d}L_{1.4}}\, \frac{L_{(1+z)\nu, i}(L_{1.4})}{4\pi} \, \left(1+\delta_{i}\right), \label{eq:jnu_radio} \\
L_{(1+z)\nu, i} =& \frac{L_{1.4}}{(1+z)^2}\left[\frac{(1+z)\nu}{1.4\, {\rm GHz}}\right]^{-\alpha_{i}}, \label{eq:radio_luminosity}
\end{align}
where $L_{1.4}$ is the radio luminosity at 1.4 GHz, and the index of $i$ runs over BL Lac objects and steep-spectrum sources\footnote{
We ignore the contributions from FSRQs in the following, since the FSRQs will have much smaller number density than other two sources
at the relevant redshift of $\simlt1$.
}.
In Eq.~(\ref{eq:jnu_radio}), $\delta_{i}$ represents the fluctuation in number density of radio sources.
In this paper, we predict the term of $\delta_{i}$ by using the following halo-occupation distribution (HOD):
\beq
\delta_{i}(\bd{x}, M) \propto \exp\left(-\frac{M_{{\rm cut}, i}}{M}\right) \delta^{(3)}_{\rm D}(\bd{x}), \label{eq:HOD_radio}
\eeq
where $\delta^{(n)}_{\rm D}$ is the $n$-dimensional Dirac delta function and 
we assume all the radio sources locate at the center of their host halo. 
The functional form of HOD is motivated by the study in \citet{2008MNRAS.391.1674W}.
For steep-spectrum radio sources, we adopt the best-fit value of $M_{\rm cut}=9.65\times10^{13}\, h^{-1}M_{\odot}$
from \citet{2008MNRAS.391.1674W}, whereas we examine various values of $M_{\rm cut}$ for BL Lac objects.
Note that the steep-spectrum sources dominate the observed flux counts at $\sim$GHz frequency in this model \citep{2010MNRAS.404..532M}.
This indicates that the clustering measurements of radio sources in the literature should be mainly determined 
by the clustering of steep-spectrum sources within our framework, allowing us to vary the typical host halo mass for BL Lac objects.

Given the model as above, we can compute the cross power spectrum of radio sources and lensing convergence as
\begin{align}
C^{\rm R}_{\kappa-T}(\nu|\ell) =& C^{\rm R, 1h}_{\kappa-T}(\nu|\ell) + C^{\rm R, 2h}_{\kappa-T}(\nu|\ell),\\
C^{\rm R, 1h}_{\kappa-T}(\nu|\ell) =& 
\sum_{i} \int_{0}^{\chi_H} \frac{{\rm d}\chi}{\chi^2}\, W_{{\rm R}, i}(\chi)\, W_{\kappa}(\chi)\, \bar{\rho}^{-1}_{\rm m} \nonumber \\
&
\times
\left[\int_{M_{\rm min}}^{M_{\rm max}}\, {\rm d}M\, \frac{{\rm d}n}{{\rm d}M}\, \exp\left(-\frac{M_{{\rm cut}, i}}{M}\right) 
\tilde{\rho}_{\rm NFW}(k=\ell/\chi, M, z)
\right] \nonumber \\
&
\times
\left[\int_{M_{\rm min}}^{M_{\rm max}}\, {\rm d}M\, \frac{{\rm d}n}{{\rm d}M}\, \exp\left(-\frac{M_{{\rm cut}, i}}{M}\right)\right]^{-1}, \\
C^{\rm R, 2h}_{\kappa-T}(\nu|\ell) =&
\sum_{i} \int_{0}^{\chi_H} \frac{{\rm d}\chi}{\chi^2}\, W_{{\rm R}, i}(\chi)\, W_{\kappa}(\chi)\, \bar{\rho}^{-1}_{\rm m}\, P_{\rm L}(k=\ell/\chi, z)\, \nonumber \\
&
\times
\left[\int_{M_{\rm min}}^{M_{\rm max}}\, {\rm d}M\, \frac{{\rm d}n}{{\rm d}M}\, \tilde{\rho}_{\rm NFW}(k=\ell/\chi, M, z) \, b(M,z)\right] 
\nonumber \\
&
\times
\left[\int_{M_{\rm min}}^{M_{\rm max}}\, {\rm d}M\, \frac{{\rm d}n}{{\rm d}M}\, \exp\left(-\frac{M_{{\rm cut}, i}}{M}\right) b(M,z)\right] \nonumber \\
&
\times
\left[\int_{M_{\rm min}}^{M_{\rm max}}\, {\rm d}M\, \frac{{\rm d}n}{{\rm d}M}\, \exp\left(-\frac{M_{{\rm cut}, i}}{M}\right)\right]^{-1},
\end{align}
where $\tilde{\rho}_{\rm NFW}$ is the Fourier transform of NFW profile and the effective window in radio $W_{{\rm R},i}$ is defined as
\begin{align}
W_{{\rm R},i}(\chi(z)) =& \frac{1}{T_0} \left(\frac{\partial B_{\nu}}{\partial T}\Bigg|_{T=T_0}\right)^{-1} \nonumber \\ 
&
\times
\frac{1}{1+z}\int_{L_{\rm min}}^{L_{\rm max}} {\rm d}L_{1.4}\, \frac{{\rm d}n_i}{{\rm d}L_{1.4}}\, \frac{L_{(1+z)\nu, i}(L_{1.4})}{4\pi},
\end{align}
where we set $L_{\rm min}=10^{30}\, {\rm erg}\, {\rm s}^{-1}{\rm Hz}^{-1}$ and $L_{\rm max}=10^{50}\, {\rm erg}\, {\rm s}^{-1}{\rm Hz}^{-1}$.

\subsubsection*{Cosmic Infrared Background}

For CIB, we follow the model developed in \citet{2012MNRAS.421.2832S}. In this model, Eq.~(\ref{eq:emission_coef}) is rewritten as
\begin{align}
j_{\nu} =&
\int {\rm d}M\, \frac{{\rm d}n}{{\rm d}M}\, \frac{1}{4\pi}\,
\Bigg[ N_{\rm cen}L_{(1+z)\nu, {\rm cen}}(M, z) \nonumber \\
&
\,\,\,\,\,
\,\,\,\,\,
\,\,\,\,\,
\,\,\,\,\,
\,\,\,\,\,
+ \int {\rm d}m\, \frac{{\rm d}n_{\rm sub}}{{\rm d}m}(m, M, z)\, L_{(1+z)\nu, {\rm sat}}(m, z)
\Bigg], \label{eq:emission_coef_CIB} \\
\equiv& \int {\rm d}M\, \frac{{\rm d}n}{{\rm d}M}\, \left[ f_{\nu, {\rm cen}}(M,z)+f_{\nu, {\rm sat}}(M,z)\right], \label{eq:emission_coef_CIB_v2}
\end{align}
where 
$N_{\rm cen}$ is the HOD of central galaxies, 
$L$ is the infrared luminosity, $m$ is the subhalo mass, and ${\rm d}n_{\rm sub}/{\rm d}m$ is the subhalo mass function.
In this paper, we use the model of subhalo mass function in \citet{2010ApJ...719...88T}.
As seen in Eq.~(\ref{eq:emission_coef_CIB}), the model assumes the statistical relation between the luminosity $L$ 
and (sub)halo mass and terms of $L_{(1+z)\nu}$ are responsible for the $L-M$ relation.
For simplicity, we assume there are no differences of the $L-M$ relation between halos and subhalos, i.e. $L_{(1+z)\nu, {\rm cen}} 
= L_{(1+z)\nu, {\rm sat}}$.
The functional form of $L_{(1+z)\nu}$ is characterized with seven physical parameters. 
We also assume $N_{\rm cen} = 1$ for $M > M_{\rm cen}$ 
and $0$ otherwise.
We adopt the best-fit parameters in the $L-M$ relation 
and the value of $M_{\rm cen}$ to the recent CIB measurement by the {\it Planck} satellite \citep{2014A&A...571A..30P}.

Hence, we can express the cross power spectrum between CIB and lensing convergence as
\begin{align}
C^{\rm CIB}_{\kappa-T}(\nu|\ell) =& C^{\rm CIB, 1h}_{\kappa-T}(\nu|\ell) + C^{\rm CIB, 2h}_{\kappa-T}(\nu|\ell),\\
C^{\rm CIB, 1h}_{\kappa-T}(\nu|\ell) =& 
\int_{0}^{\chi_H} \frac{{\rm d}\chi}{\chi^2}\, W_{\rm CIB}(\chi) \, W_{\kappa}(\chi)\, 
\bar{\rho}^{-1}_{\rm m} \nonumber \\
&
\times
\int_{M_{\rm min}}^{M_{\rm max}}\, {\rm d}M\, \frac{{\rm d}n}{{\rm d}M}\, 
\Bigg[f_{\nu, {\rm cen}}(M,z) \nonumber \\
&
+
f_{\nu, {\rm sat}}(M,z)\,\tilde{u}_{\rm sat}(k=\ell/\chi, M, z)\Bigg]\tilde{\rho}_{\rm NFW}(k=\ell/\chi, M, z), \\
C^{\rm CIB, 2h}_{\kappa-T}(\nu|\ell) =&
\int_{0}^{\chi_H} \frac{{\rm d}\chi}{\chi^2}\, W_{\rm CIB}(\chi) \, W_{\kappa}(\chi)\, \bar{\rho}^{-1}_{\rm m}\, P_{\rm L}(k=\ell/\chi, z)\, \nonumber \\
&
\times
\left[\int_{M_{\rm min}}^{M_{\rm max}}\, {\rm d}M\, \frac{{\rm d}n}{{\rm d}M}\, \tilde{\rho}_{\rm NFW}(k=\ell/\chi, M, z) \, b(M,z)\right] 
\nonumber \\
&
\times
\Bigg[\int_{M_{\rm min}}^{M_{\rm max}}\, {\rm d}M\, \frac{{\rm d}n}{{\rm d}M}\, \Big( f_{\nu, {\rm cen}}(M,z)
\nonumber \\
&
\,\,\,\,\,
+f_{\nu, {\rm sat}}(M,z) \tilde{u}_{\rm sat}(k=\ell/\chi, M, z) \Big)
b(M,z)\Bigg],
\end{align}
where $u_{\rm sat}(\bd{x})$ is the number density profile of satellite galaxies normalized to $\int u_{\rm sat}\, {\rm d}V =1$,
and $\tilde{u}_{\rm sat}$ is the Fourier counterpart. We assume $u_{\rm sat}=\rho_{\rm NFW}/M$ throughout this paper.
The kernel function of $W_{\rm CIB}(\chi)$ is given by
\beq
W_{\rm CIB}(\chi(z)) = \frac{1}{T_0} \left(\frac{\partial B_{\nu}}{\partial T}\Bigg|_{T=T_0}\right)^{-1} \frac{1}{1+z}.
\eeq

\subsection{Effective redshifts in cross correlations of astrophysical sources and lensing}

Before detailed computations, we shall show the effective redshift range to be probed 
by the cross correlation between astrophysical sources at millimeter wavelengths and lensing convergence.
For this purpose, we compute the mean intensity from cumulative emission from radio sources and CIB.
For $i$-th radio source (BL Lac object or steep-spectrum source), the mean intensity is given by
\beq
I^{\rm R}_{\nu, i} = \int_{0}^{\chi_H}\frac{{\rm d}\chi}{1+z}\,\int_{L_{\rm min}}^{L_{\rm max}} {\rm d}L_{1.4}\, \frac{{\rm d}n_i}{{\rm d}L_{1.4}}\, \frac{L_{(1+z)\nu, i}(L_{1.4})}{4\pi},
\eeq
while the CIB mean intensity can be computed as
\beq
I^{\rm CIB}_{\nu} = \int_{0}^{\chi_H}\frac{{\rm d}\chi}{1+z}\, \int_{M_{\rm min}}^{M_{\rm max}}\, {\rm d}M\, \frac{{\rm d}n}{{\rm d}M}\, \left[ f_{\nu, {\rm cen}}(M,z)+f_{\nu, {\rm sat}}(M,z) \right].
\eeq

\begin{figure}
\centering
\includegraphics[width=0.80\columnwidth]
{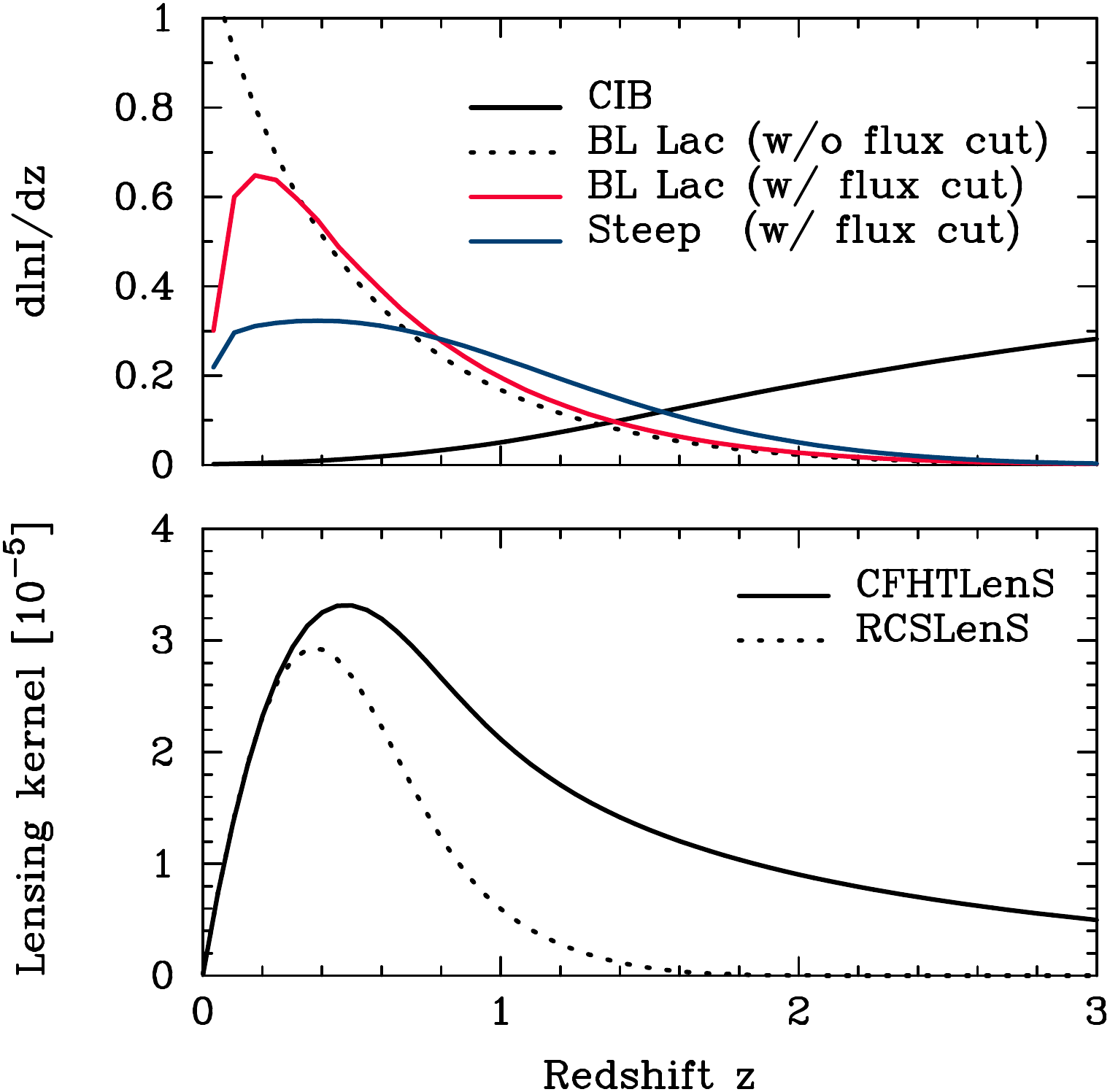}
\caption{
{\it Top}: Redshift dependence of mean intensity from cosmic infrared background (CIB)
and cumulative radio sources. We show the CIB intensity at 217 GHz and radio intensity at 100 GHz.
{\it Bottom}: The lensing efficiency in current lensing surveys.
}
\label{fig:dlnIdz_Wkappa}
\end{figure}

The top panel in figure~\ref{fig:dlnIdz_Wkappa} shows the redshift dependence in $I^{\rm R}_{\nu, i}$ at 100 GHz and $I^{\rm CIB}_{\nu}$
at 217 GHz and the bottom represents the lensing kernel $W_{\kappa}(\chi)$ 
for two lensing surveys of CFHTLenS \citep{2013MNRAS.433.3373V} and RCSLenS \citep{2017MNRAS.471.1565H}. In the top panel, the solid line shows the CIB intensity, while the red and blue lines are the intensity coming from BL Lac objects and steep-spectrum sources, respectively.
To compute the colored lines, we assume the flux cut of $S_{\rm lim}=400\, {\rm mJy}$ at 100 GHz which is taken from Table 1 in \citet{2014A&A...571A..30P}.
First of all, the main contribution in CIB intensity will come from star-forming galaxies at higher redshift of $\simgt2$, making the cross correlation with galaxy lensing irrelevant. 
In contrast, the radio sources can have a sizable cross correlation with large-scale structures at $z\sim0.2-0.4$.
Interestingly, BL Lac objects, minor population in radio flux counts at $\sim1$GHz, 
can dominate a possible correlation with gravitational lensing in imaging survey.
The model predicts BL Lac objects can be the main contributor to the mean intensity 
at lower redshifts ($z<1$) because of the flatness of their spectral index.
We also see the impact of flux cut by comparing the red line with dashed line in the top panel in figure~\ref{fig:dlnIdz_Wkappa}.
The flux cut in the {\it Planck} satellite can remove the BL Lac objects at $z\simlt0.1$, but the objects at $z\sim0.2-0.4$ will still survive
in the observed temperature maps.

\begin{figure}
\centering
\includegraphics[width=0.80\columnwidth]
{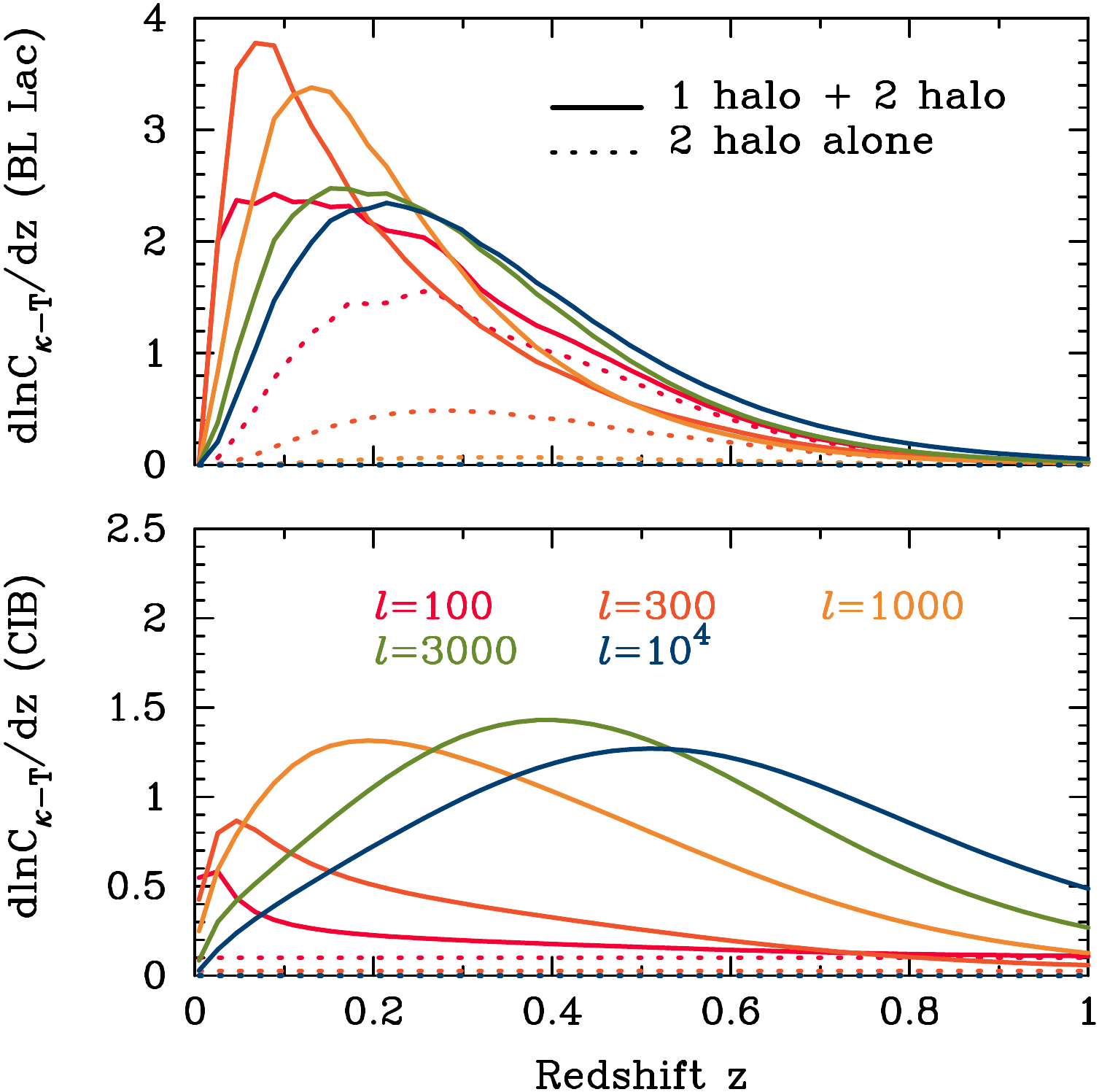}
\caption{
\ms{The contribution of large-scale structure at different redshifts to the cross power spectrum of weak lensing
and brightness temperature. In each panel, the solid lines show the sum of one-halo and two-halo terms for a fixed multipole,
while the dashed lines are for two-halo term alone. The different colored lines represent the difference in multipole $\ell$.
In this figure, we adopt the source redshift distribution in CFHTLenS \citep{2013MNRAS.433.3373V}.
{\it Top}: The cumulative emission of flat-spectrum radio sources at 100 GHz. We assume the HOD parameter $M_{\rm cut} = 9.65\times10^{13}\,h^{-1}M_{\odot}$. {\it Bottom}: The CIB intensity at 217 GHz.}
}
\label{fig:dlnCdz}
\end{figure} 

\ms{
Furthermore, Figure~\ref{fig:dlnCdz} represents the redshift derivative of the cross power spectrum ${\rm d}\ln C_{\kappa-T}/{\rm d}z$,
showing which redshift sources the CIB or radio contamination to the tSZ-lensing signal as a function of angular scale.
In this figure, the solid line shows the contribution from both of one-halo and two-halo terms, while the dashed lines are for two-halo term alone.
For the radio source, we find the effective redshift to be $0.2-0.4$ over the wide range of $\ell=100-10^{4}$ since the lensing kernel has a similar redshift dependence of the radio intensity.
On the other hand, the CIB-lensing correlation would arise from large-scale structure at various redshifts, 
and the correlation on smaller angular scale can be determined by the structure at higher redshift.
}

\section{Results}
\label{sec:result}

\subsection{Frequency dependence on cross correlations of astrophysical sources and lensing}
\label{subsec:nu_depend}

\begin{figure}
\centering
\includegraphics[width=0.80\columnwidth]
{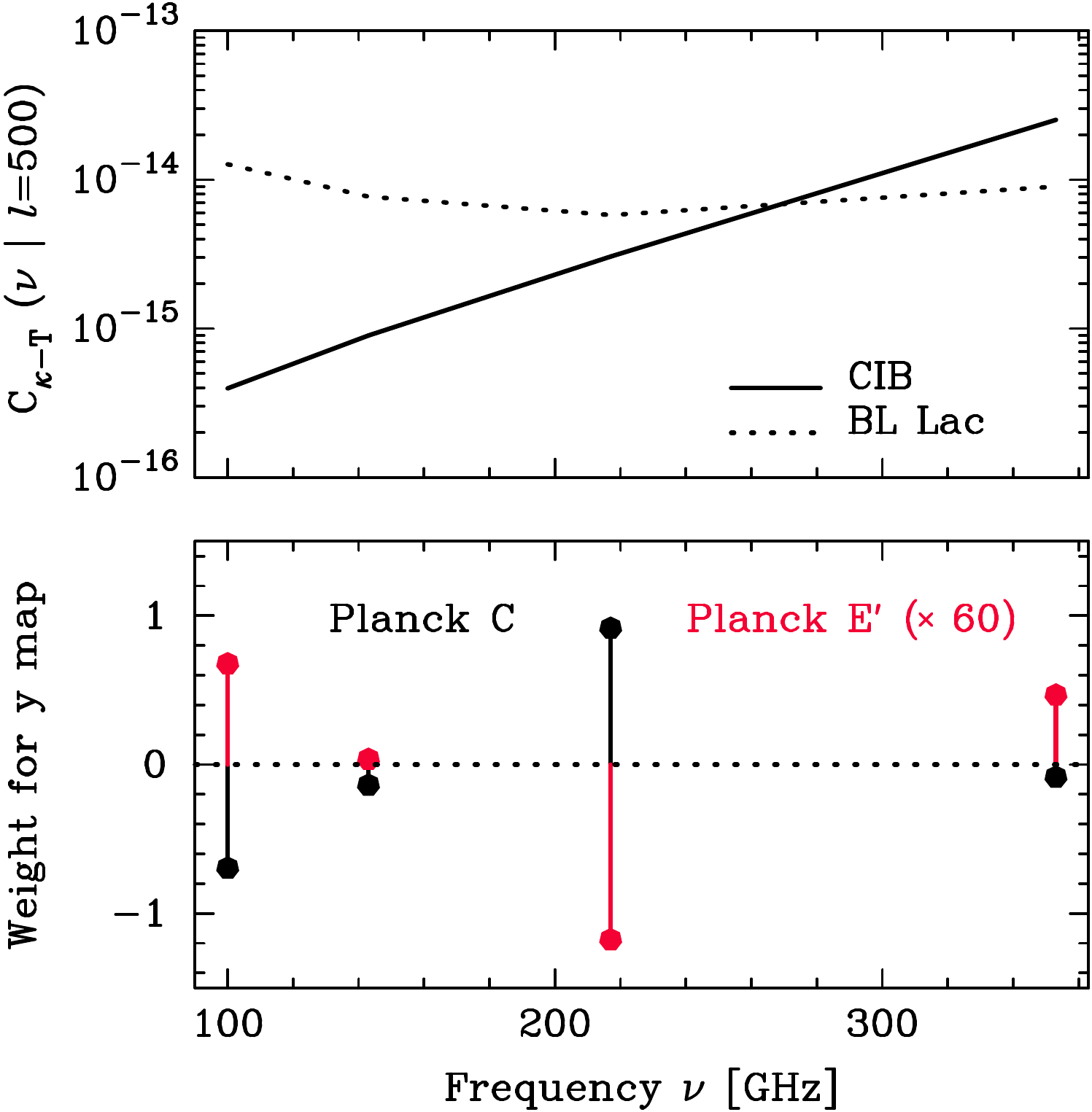}
\caption{
{\it Top}: Our model of cross power spectrum between brightness temperature in {\it Planck} bands and weak lensing as a function of frequency. 
Solid line in the top panel is the cross power spectrum between CIB and galaxy weak lensing at multipole $\ell=500$, 
while dashed line shows the cross power spectrum between flat-spectrum radio source and lensing.
We assume the source redshift distribution of galaxy lensing measurement in CFHTLenS \citep{2014PhRvD..89b3508V}.
{\it Bottom}: Relative contribution from multiple frequencies in construction of Compton $y$ map.
Black lines show the weight for construction of Compton $y$ map used for measurement of tSZ-lensing correlation, 
while red represents the weight for evaluation of residual foreground contamination in tSZ-lensing correlation \citep{2014PhRvD..89b3508V}.
}
\label{fig:Cl_weight_vs_nu}
\end{figure}

The frequency dependence on the cross power spectrum $C_{\kappa-T}$ in Eq.~(\ref{eq:obs_yk_power})
is the key to understanding the observed tSZ-lensing correlation within our framework.
Figure~\ref{fig:Cl_weight_vs_nu} shows the cross spectrum $C_{\kappa-T}$ at $\ell=500$ as a function of frequency.
Note that $\ell=500$ roughly corresponds to 6-7 arcmin in angular scale and it is 
relevant for the tension between the observed $y-\kappa$ correlation 
and an expected signal from tSZ effect in ICM \citep{2015JCAP...09..046M}.
In the top panel, solid line represents the CIB-lensing power spectrum, 
while dashed line is for the radio-lensing power spectrum from BL Lac objects.
We here assume the HOD parameter $M_{\rm cut} = 9.65\times10^{13}\,h^{-1}M_{\odot}$ for BL Lac objects as same as steep-spectrum sources.  
In the bottom, we also plot the weight for construction of Compton $y$ map.
Black lines in the bottom correspond to the weights for extracting tSZ effect from observed temperature maps,
while red line is the specific case so as to remove tSZ effects in observed Compton $y$ map.
As shown in the figure, the CIB-lensing correlation has a steep spectrum and it becomes important
if high-frequency maps are weighted for $y$-map construction.
On the other hand, the correlation with BL Lac objects and lensing convergence shows almost flat spectrum 
but it will be likely to dominate the cross correlation at $\sim$100 GHz.
Since one need a negative weight at 100 GHz with a large amplitude to obtain unbiased estimate of tSZ effect in practice (see the black line in
the bottom), the radio-lensing correlation can induce a {\it negative} correction for observed tSZ-lensing correlation\footnote{Note that \citet{2014A&A...571A..21P, 2014A&A...571A..29P} have also discussed similar effects; the negative response of the $y$-map weights to radio sources and the positive response due to dust emission.}.
In addition, we expect a non-zero cross correlation arising from various terms in $C_{\kappa-T}$ even 
if working with the weight to be $\sum_{i} w_{i} g(x_i) = 0$ as in red in the bottom panel.

\subsection{Comparison
of observed tSZ-lensing cross correlation with our model}

Let us then make a comparison of the observed tSZ-lensing cross correlations \citep{2014PhRvD..89b3508V, 2017MNRAS.471.1565H} with our model prediction.
When predicting the $y-\kappa$ correlation in CFHTLenS, we set the gaussian filter 
with the beaming size of 6 arcmin for lensing convergence and the FWHM of 9.5 arcmin for Compton $y$ map \citep{2014PhRvD..89b3508V}.
Similarly, we adopt the gaussian filter with the FWHM of 10.0 arcmin for both of $y$ and $\kappa$ maps in RCSLenS \citep{2017MNRAS.471.1565H}.
Note that we apply the weight defined in CFHTLenS analyses for the RCSLenS predicition. This simplified procedure 
can induce a $10\%$-level uncertainty in theoretical model, while it is less problematic under the current statistical uncertainty \citep{2017MNRAS.471.1565H}.

\begin{figure}
\includegraphics[width=0.99\columnwidth]
{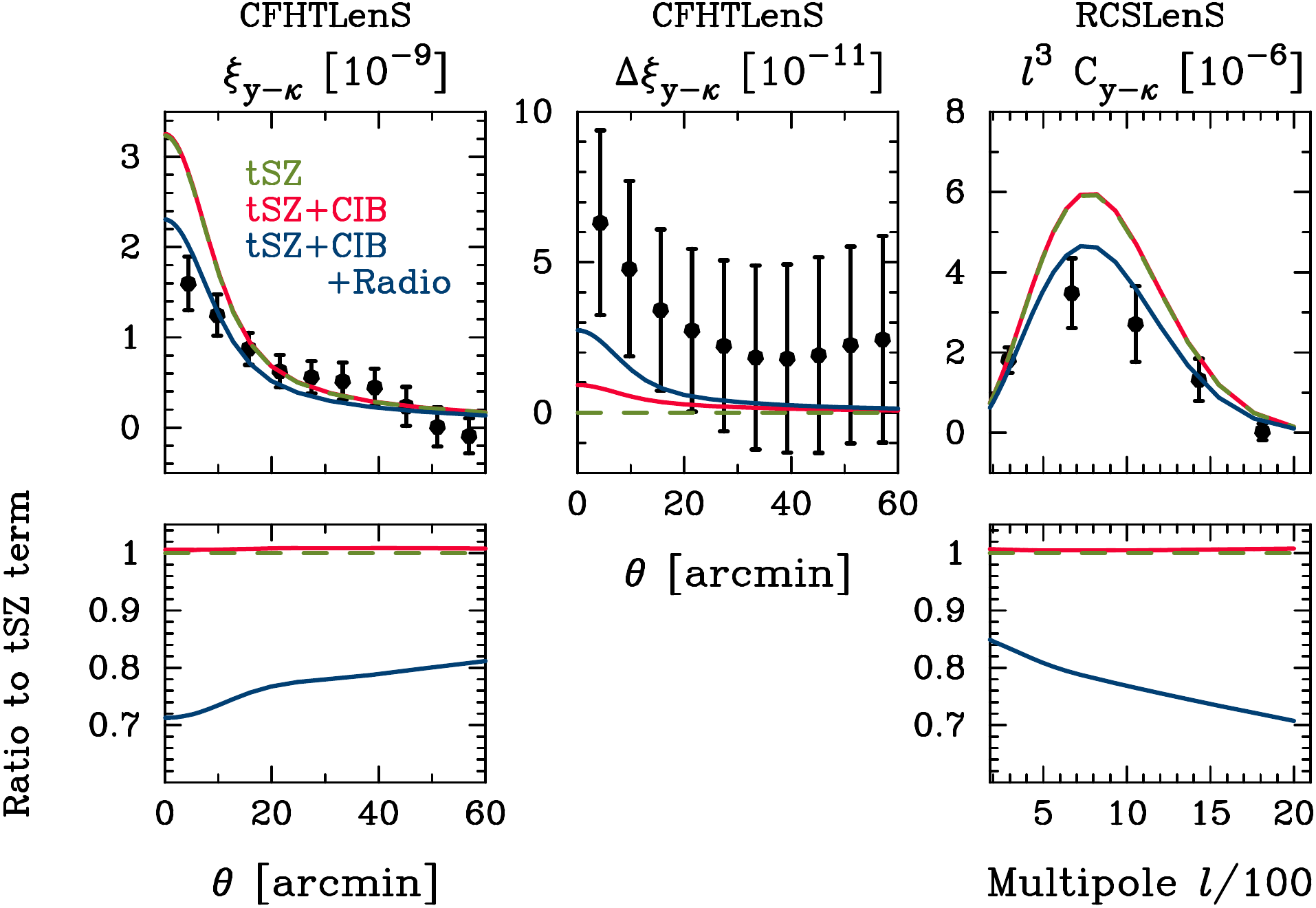}
\caption{
Comparison of $y-\kappa$ correlation with recent measurements and our prediction.
In the top three panels, the black points with error bar show the measurements of $y-\kappa$ correlation 
in CFHTLenS \citep{2014PhRvD..89b3508V} and RCSLenS \citep{2017MNRAS.471.1565H}. 
Top left and right panels show the $y-\kappa$ correlations which are expected 
to be dominated in tSZ-lensing correlation, while top middle panel represents the $y-\kappa$ correlation in the absence of tSZ effects.
In top panels, green dashed line shows the expected correlation from tSZ effects by ICM, red lines include the residual correlation from CIB and blue lines contain the residuals from CIB and radio sources. In the bottom panels, we show the ratio of 
the $y-\kappa$ correlations with residuals from CIB and/or radio sources to the tSZ$-\kappa$ correlation expected from ICM.
In this figure, we assume that the flat-spectrum radio sources live in a massive host halo with the HOD parameter of $M_{\rm cut} = 10^{15}\, h^{-1}\, M_{\odot}$.
}
\label{fig:yk_correlation_best}
\end{figure} 

\begin{figure}
\includegraphics[width=0.99\columnwidth]
{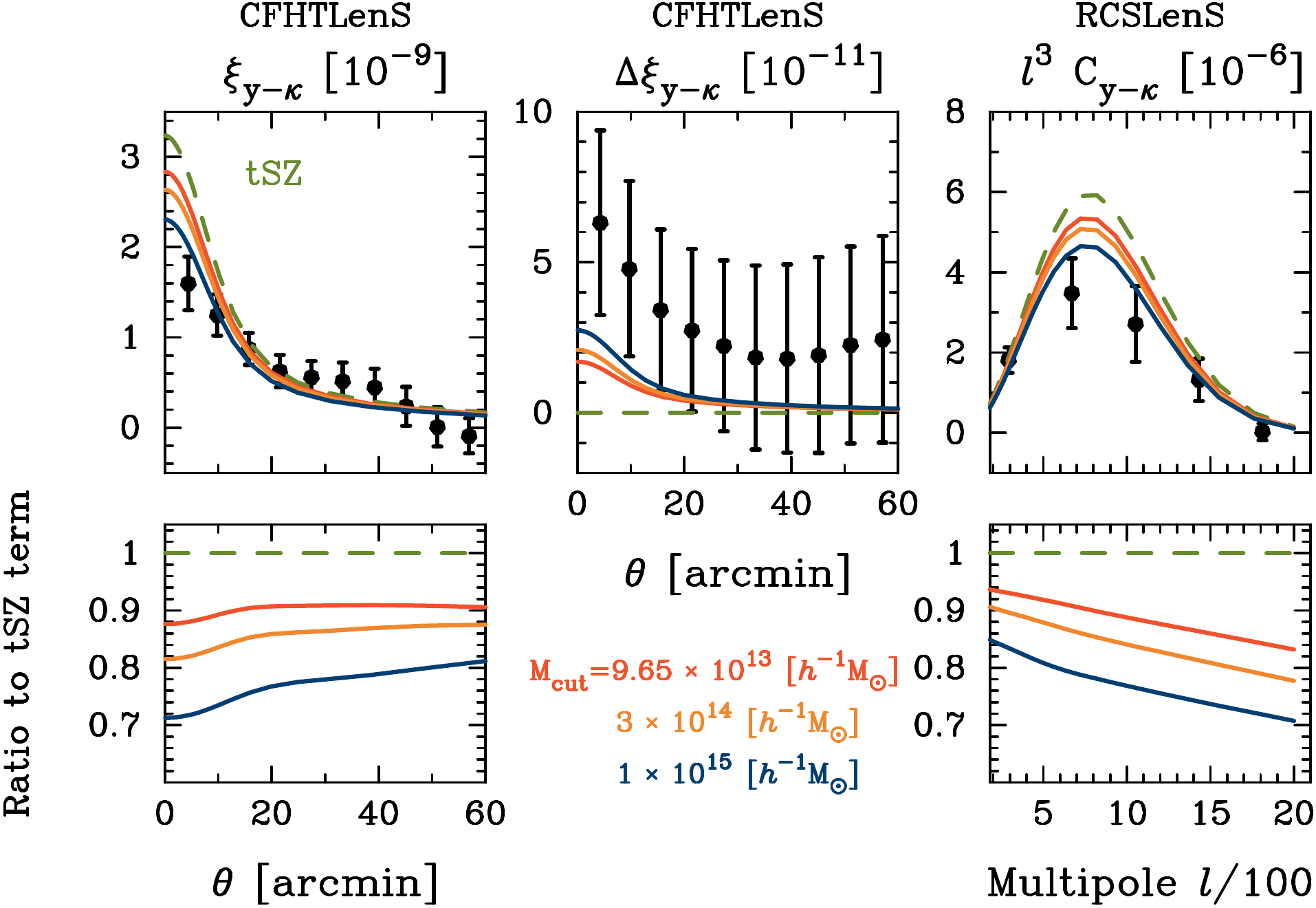}
\caption{
Similar to Figure~\ref{fig:yk_correlation_best}, but this shows the dependence of 
the mass parameter in HOD of flat-spectrum radio sources, denoted by $M_{\rm cut}$.
}
\label{fig:yk_correlation_Mcut}
\end{figure}

Figure~\ref{fig:yk_correlation_best} summarizes the comparison with the observed correlation and our model prediction.
The black points with error bar in the top panels show the observed correlation taken from \citet{2014PhRvD..89b3508V} and \citet{2017MNRAS.471.1565H}.
In the top panels, green dashed lines represent the expected correlation coming from tSZ effect in ICM 
(see Section~\ref{subsec:ICM}), while red and blue lines include the corrections induced by the cross correlation of lensing
and astrophysical sources. The red lines correspond to the signal including the CIB-$\kappa$ correlation, showing the CIB
plays a minor role in observed $y-\kappa$ correlation with weights for construction of unbiased tSZ effect (i.e. the case of $\sum_i w_i g(x_i) =1$).
Including the radio sources can decrease the observed $y-\kappa$ correlation as discussed in Section~\ref{subsec:nu_depend}.
We find that the model with $M_{\rm cut}=10^{15}\, h^{-1}M_{\odot}$ for BL Lac objects can provide a reasonable fit 
to the observed $y-\kappa$ correlation in both of CFHTLenS and RCSLenS.
\ms{Note that we change single parameter of $M_{\rm cut}$ in the HOD of flat-spectrum radio sources by hand and
keep other parameters fixed to find the model in Figure~\ref{fig:yk_correlation_best}.}
This model can also explain the correlation with lensing convergence and Compton $y$ map 
in the absence of tSZ effect as shown in the top middle panel.
The relative correction for ICM-lensing correlation is found 
to be about a $\sim2\%$ level from CIB, while the radio sources can induce a $20-30\%$ level correction. 
For a conservative scenario setting the same $M_{\rm cut}$ in BL Lac objects as in steep-spectrum sources, 
we need to take into account the correction with a level of $\sim10\%$ for observed $y-\kappa$ correlation.
It would be worth noting that the conservative scenario looks inconsistent with the observed $y-\kappa$ correlation
in the absence of tSZ effect, but the model with $M_{\rm cut}=10^{15}\, h^{-1}M_{\odot}$ is in better agreement with (see Figure~\ref{fig:yk_correlation_Mcut})\footnote{
To place a meaningful constraint of $M_{\rm cut}$, 
the current measurements would be insufficient due to the degeneracy among parameters in our model.
To break the degeneracy among cosmology, the ICM, and residual components from CIB and faint radio sources,
the tomographic tSZ-lensing correlation is expected to be essential. We will work on it in the near future.
}.

Can the radio sources be allowed to live in massive cluster-sized halos with masses of $\sim10^{15}\, h^{-1}M_{\odot}$ 
at redshift of $z=0.2-0.4$? This is still an open question when we think of flat-spectrum radio sources like BL Lac objects,
since the measurements of clustering of radio sources have been performed at frequency of $\sim$GHz 
where steep-spectrum sources should be abundant.
When adopting the model with $M_{\rm cut}=10^{15}\, h^{-1}M_{\odot}$, we can infer the expected halo bias of flat-spectrum sources
to be $\sim5-8$ at $0<z<1$. This is found to be significantly higher than the halo bias of radio galaxies \citep[e.g.][]{2014MNRAS.440.1527L, 2014MNRAS.440.2322L, 2015ApJ...812...85N, 2015MNRAS.451..849A} and radio-loud AGNs \citep[e.g.][]{2009ApJ...697.1656S, 2017A&A...600A..97R}. This finding is still in no contradiction to the existing clustering analyses of radio sources,
since one can determine the halo bias of steep-spectrum sources alone from the clustering analyses in the literature.

\section{Conclusion}
\label{sec:conc}

In this paper, we studied the impact of astrophysical sources at multiple frequencies in the CMB measurement
on the cross correlation between thermal Sunyaev-Zel'dovich (tSZ) effect and weak gravitational lensing effect.
We developed a halo model to predict possible correlations between astrophysical sources and lensing convergence $\kappa$.
Starting from an estimator of Compton $y$ map in the CMB measurement, we found the correlation between astrophysical sources
and lensing effect in galaxy imaging survey may affect the observed $y-\kappa$ correlation.
Assuming the best-fit model of cosmic infrared background (CIB) to the recent observation, we evaluated 
the CIB-lensing correlation is less important for observed $y-\kappa$ correlation in current imaging surveys.
In contrast, the radio sources with flat-spectral index of $\sim0.1$ can be the main contributor to 
the observed extragalctic intensity at $\sim100$GHz and the cross correlation between such flat-spectrum sources and lensing
induces a negative correction for the observed $y-\kappa$ correlation with a level of 20-30\% 
if the flat-spectrum sources could populate most massive dark matter halos with masses of $\sim10^{15}\, h^{-1}M_{\odot}$. 
Including possible negative corrections from radio-lensing correlation enables us to explain the observed $y-\kappa$ correlation 
without introducing strong AGN feedback nor a small amplitude in linear matter density fluctuations.

A caveat in our model is that the model relies on the cross correlation between lensing convergence and a minor population in radio bands.
\ms{In fact, there are still large theoretical uncertainties in the modeling of the flat-spectrum radio sources (in particular of their HOD).
To improve our understanding of the observed tSZ-lensing correlation, we require additional observational tests to
learn about the radio sources at $\sim100$ GHz.}
The previous measurements of clustering in radio galaxies and AGNs would not be helpful 
to improve our understanding of the radio-lensing correlation by flat-spectrum sources,
since most of clustering measurements are subject to another population called steep-spectrum sources.
We expect that the measurement of cross correlation of galaxy lensing with Compton $y$ map {\it in the absence of tSZ effect} (i.e. $\sum_i w_i g(x_i) =0$ in Eq.~\ref{eq:ILC})
is a possible approach to determine the statistical relationship between faint flat-spectrum radio sources and dark matter halos
\citep[see also][for the study of clustering of BL Lac objects at gamma-ray frequencies]{PhysRevD.97.123015}.
Such a measurement has large statistical uncertainty at present, whereas ongoing and future imaging surveys
can change the current situation, allowing to establish a precise theoretical framework to relate flat-spectrum radio sources with large-scale structures. \ms{It would be worth noting that future studies should investigate optimal weighting of temperature maps over different frequencies
to search for the relationship between radio sources and their host halos.}
The clustering analyses of flux-limited sample at $\sim100\,$GHz should be a complementary approach, 
but the number density of flat-spectrum sources is evaluated to be of an order of $\sim100\,{\rm str}^{-1}$ with 
the flux cut of $400$mJy at 100 GHz. Deeper observations at $\sim100\,{\rm GHz}$ will enable us to study the HOD of flat-spectrum sources in detail
by increasing their number density.

It is also important for precise measurement of $y-\kappa$ cross-correlation in the future to consider 
other systematic effects, 
for instance, the correlation between intrinsic alignment of satellite galaxies in galaxy clusters \citep[e.g.][]{2015PhR...558....1T}
and the tSZ effect from the same clusters.
Furthermore, it is worth exploring if lensing tomography can 
mitigate the impact of radio sources on observed $y-\kappa$ correlation
and developing some approach to have an unbiased estimate of tSZ-lensing correlation by fully utilizing the frequency dependence on radio-lensing correlation.
We will leave those for our future studies.

\section*{Acknowledgements}
The author appreciates careful reading and suggestion to improve the article by anonymous referees.
The author thanks Naoki Yoshida and J.~Colin Hill for helpful comments.
This work was in part supported by 
Grant-in-Aid for Scientific Research on Innovative Areas
from the MEXT KAKENHI Grant Number (18H04358).  
Numerical computations presented in this paper were in part carried out 
on the general-purpose PC farm at Center for Computational Astrophysics, CfCA, 
of National Astronomical Observatory of Japan.


\appendix

\section{Halo model of the cross correlation between weak lensing convergence and astrophysical sources at CMB frequencies}

In this appendix, we formulate the cross correlation function between weak lensing convergence and astrophysical sources based on halo-model approach \citep{2002PhR...372....1C}. The lensing convergence $\kappa$ and cumulative emission from astrophysical sources at a given frequency $\nu$
are expressed as
\begin{align}
\kappa(\bd{\theta}) = & \int\, {\rm d}\chi\, W_{\kappa}(\chi)\, \delta_{\rm m}(r(\chi)\bd{\theta}, z(\chi)), \\
I_{\nu}(\bd{\theta}) = & \int\, {\rm d}\chi\, W_{I}(\chi)\, j_{\nu}(r(\chi)\bd{\theta}, z(\chi)),
\end{align}
where $I_{\nu}$ is the observed specific intensity, $j_{\nu}$ is the comoving specific emission coefficient, 
$W_{I}(\chi) = [1+z(\chi)]^{-1}$, and the kernel of $W_{\kappa}$ is given by Eq.~(\ref{eq:lens_kernel}).
Using the Limber approximation \citep{1954ApJ...119..655L}, we can write the cross power spectrum between $\kappa$ and $I_{\nu}$ as
\begin{align}
C_{\kappa-I}(\nu|\ell) = \int \frac{{\rm d}\chi}{\chi^2} W_{\kappa}(\chi)\, W_{\nu}(\chi)\, P_{{\rm m}-j\nu}\left(k=\frac{\ell}{\chi}, z(\chi)\right),
\end{align}
where $P_{{\rm m}-j{\nu}}(k, z)$ is the three-dimensional cross power spectrum between matter overdensity $\delta_{\rm m}(\bd{x})$
and the comoving specific emission coefficient $j_{\nu}(\bd{x})$ at redshift of $z$.

In the halo model, underlying matter density field at a given redshift $z$ can be approximated as
\begin{align}
\rho_{\rm m}(\bd{x}) =& \sum_{i} \rho_{\rm h}(\bd{x}-\bd{x}_{i}|M_{i},z) \nonumber \\
=& \sum_{i} \int {\rm d}M\, \delta_{\rm D}(M-M_{i}) \int {\rm d}^3{\bd x}^{\prime}\, \delta^{(3)}_{D}(\bd{x}^{\prime}-\bd{x}_{i})\, \rho_{\rm h}(\bd{x}-\bd{x}^{\prime}|M,z), \label{eq:rhom_halomodel}
\end{align}
where $\rho_{\rm h}$ is the density profile of a dark matter halo
and $\delta^{(n)}_{\rm D}(\bd{x})$ is the $n$-dimensional Dirac delta function.
In the following equations, we omit the redshift $z$ for simplicity.
Within the halo-model framework, the halo mass function and two-point correlation of halos are defined as
\begin{align}
&\left\langle \sum_{i} \delta_{\rm D}(M-M_{i}) \delta^{(3)}_{D}(\bd{x}-\bd{x}_{i}) \right\rangle \equiv \frac{{\rm d}n}{{\rm d}M}, \label{eq:mf_halomodel}\\
&\left\langle \sum_{i,j|i\neq j} \delta_{\rm D}(M-M_{i}) \delta^{(3)}_{D}(\bd{x}_{1}-\bd{x}_{i}) \delta_{\rm D}(M^{\prime}-M_{j}) \delta^{(3)}_{D}(\bd{x}_{2}-\bd{x}_{j}) \right\rangle \nonumber \\
&
\,\,\,\,\,\,\,
\,\,\,\,\,\,\,
\,\,\,\,\,\,\,
\,\,\,\,\,\,\,
\,\,\,\,\,\,\,
\,\,\,\,\,\,\,
\,\,\,\,\,\,\,
\,\,\,\,\,\,\,
\equiv  \frac{{\rm d}n}{{\rm d}M}\frac{{\rm d}n}{{\rm d}M^{\prime}}\, \xi_{\rm hh}(\bd{x}_{1}-\bd{x}_{2}|M,M^{\prime}), \label{eq:xihh_halomodel}
\end{align}
where ${\rm d}n/{\rm d}M$ is the halo mass function and 
$\xi_{\rm hh}(\bd{x} | M, M^{\prime})$ is the halo-halo correlation function with mass of $M$ and $M^{\prime}$.

Similar to Eq.~(\ref{eq:rhom_halomodel}), we can express the comoving specific emission coefficient as
\begin{align}
j_{\nu}(\bd{x}) =& \sum_{i} {\cal J}_{{\rm h},\nu}(\bd{x}-\bd{x}_{i}|M_{i}) \nonumber \\
=& \sum_{i} \int {\rm d}M\, \delta_{\rm D}(M-M_{i}) \int {\rm d}^3{\bd x}^{\prime}\, \delta^{(3)}_{D}(\bd{x}^{\prime}-\bd{x}_{i})\, {\cal J}_{{\rm h},\nu}(\bd{x}-\bd{x}^{\prime}|M), \label{eq:jnu_halomodel}
\end{align}
where ${\cal J}_{{\rm h},\nu}$ is the emission coefficient profile in single halo and we assume it depends on halo mass $M$.

We then consider the three-dimensional correlation function of $\rho_{\rm m}$ and $j_{\nu}$.
The correlation function is defined as
\beq
\bar{\rho}_{\rm m} \, \xi_{{\rm m}-j\nu}(\bd{x}_{1}-\bd{x}_2) \equiv \langle \rho_{\rm m}(\bd{x}_1) j_{\nu}(\bd{x}_2) \rangle - \bar{\rho}_{\rm m}\bar{j}_{\nu}, \label{eq:cross_xi_m_j_3d}
\eeq
where $\bar{\rho}_{\rm m}$ is the mean matter density and $\bar{j}_{\nu}$ is the spatially-averaged emission coefficient.
Using Eqs.~(\ref{eq:rhom_halomodel})--(\ref{eq:jnu_halomodel}), one can decompose $\xi_{{\rm m}-j\nu}$ into two parts:
one is so-called one-halo term arising from the correlation in single halos and another is two-halo term describing the correlation
due to clustering of two distinct halos.
The one-halo term of Eq.~(\ref{eq:cross_xi_m_j_3d}) is given by
\beq
\int {\rm d}M \, \frac{{\rm d}n}{{\rm d}M}\, \int {\rm d}^{3}\bd{y}\, 
\rho_{\rm h}(\bd{x}_{1}-\bd{y}|M) {\cal J}_{{\rm h},\nu}(\bd{x}_{2}-\bd{y}|M),
\eeq
while the two-halo term is 
\begin{align}
&\int {\rm d}M \, \frac{{\rm d}n}{{\rm d}M} b(M)\, \int {\rm d}M^{\prime} \, \frac{{\rm d}n}{{\rm d}M^{\prime}} b(M^{\prime})
\int {\rm d}^{3}\bd{y}\, \rho_{\rm h}(\bd{x}_{1}-\bd{y}|M) \nonumber \\
&
\,\,\,\,\,\,\,
\,\,\,\,\,\,\,
\,\,\,\,\,\,\,
\,\,\,\,\,\,\,
\times
\int {\rm d}^{3}\bd{y}^{\prime}\, {\cal J}_{{\rm h},\nu}(\bd{x}_{2}-\bd{y}^{\prime}|M)
\, \xi_{\rm L}(\bd{y}-\bd{y}^{\prime}), 
\label{eq:twoh_details}
\end{align}
where $b(M)$ is the linear halo bias and $\xi_{\rm L}$ is the linear matter correlation function.
In Eq.~(\ref{eq:twoh_details}), we assume $\xi_{\rm hh}(\bd{x}|M,M^{\prime}) = b(M)\, b(M^{\prime})\, \xi_{\rm L}(\bd{x})$.
Hence, the final expression of Eq.~(\ref{eq:cross_xi_m_j_3d}) is given by
\begin{align}
\xi_{{\rm m}-j\nu}(\bd{x}_{1}-\bd{x}_2)
=& \, \xi^{\rm 1h}_{{\rm m}-j\nu}(\bd{x}_{1}-\bd{x}_2) + \xi^{\rm 2h}_{{\rm m}-j\nu}(\bd{x}_{1}-\bd{x}_2), \label{eq:cc_halomodel} \\
\xi^{\rm 1h}_{{\rm m}-j\nu}(\bd{x}_{1}-\bd{x}_2) =& \int {\rm d}M \, \frac{{\rm d}n}{{\rm d}M}\, \int {\rm d}^{3}\bd{y}\,
\frac{\rho_{\rm h}(\bd{x}_{1}-\bd{y}|M)}{\bar{\rho}_{\rm m}} {\cal J}_{{\rm h},\nu}(\bd{x}_{2}-\bd{y}|M) \\
\xi^{\rm 2h}_{{\rm m}-j\nu}(\bd{x}_{1}-\bd{x}_2) =&
\int {\rm d}M \, \frac{{\rm d}n}{{\rm d}M} b(M) \int {\rm d}M^{\prime} \, \frac{{\rm d}n}{{\rm d}M^{\prime}} b(M^{\prime}) 
\nonumber \\
&
\,\,\,\,\,\,\,
\times \int {\rm d}^{3}\bd{y} \int {\rm d}^{3}\bd{y}^{\prime}\, \frac{\rho_{\rm h}(\bd{x}_{1}-\bd{y}|M)}{\bar{\rho}_{\rm m}} \nonumber \\
&
\,\,\,\,\,\,\,
\,\,\,\,\,\,\,
\,\,\,\,\,\,\,
\times
{\cal J}_{{\rm h},\nu}(\bd{x}_{2}-\bd{y}|M)\, \xi_{\rm L}(\bd{y}-\bd{y}^{\prime}). 
\end{align}
The Fourier transform of Eq.~(\ref{eq:cc_halomodel}) is the three-dimensional power spectrum of $P_{{\rm m}-j\nu}$, which is expressed as
\begin{align}
P_{{\rm m}-j\nu}(k) =& P^{\rm 1h}_{{\rm m}-j\nu}(k) + P^{\rm 2h}_{{\rm m}-j\nu}(k), \\
P^{\rm 1h}_{{\rm m}-j\nu}(k) =& \int {\rm d}M \, \frac{{\rm d}n}{{\rm d}M}\, 
\frac{\tilde{\rho}_{\rm h}(k|M)}{\bar{\rho}_{\rm m}} \tilde{{\cal J}}_{{\rm h},\nu}(k|M), \\
P^{\rm 2h}_{{\rm m}-j\nu}(k) =& 
\left[\int {\rm d}M \, \frac{{\rm d}n}{{\rm d}M} b(M) \frac{\tilde{\rho}_{\rm h}(k|M)}{\bar{\rho}_{\rm m}} \right] \nonumber \\
&
\,\,\,\,\,\,\,
\,\,\,\,\,\,\,
\times
\left[\int {\rm d}M \, \frac{{\rm d}n}{{\rm d}M} b(M) \tilde{{\cal J}}_{{\rm h},\nu}(k|M)\right]\, P_{\rm L}(k),
\end{align}
where 
$P_{\rm L}(k)$ is the linear matter power spectrum,
$\tilde{\rho}_{\rm h}$ and $\tilde{{\cal J}}_{{\rm h},\nu}$ are the Fourier counterparts of $\rho_{\rm h}$ and ${\cal J}_{{\rm h},\nu}$, respectively.

To obtain the expressions in Section~\ref{subsec:astro}, we assume radio sources to be point sources
and their halo occupation distribution (HOD) is expressed as an exponential form (see Eq.~\ref{eq:HOD_radio}).
We also assume the HOD is independent of redshift and radio luminosity. In this case, 
the emission coefficient profile of radio sources is given by
\begin{align}
&\tilde{\cal J}_{\nu}(k|M) = \bar{\cal J}_{\nu,\rm R} \, S(M), \\
&\bar{\cal J}_{\nu,\rm R} = \int {\rm d}L_{1.4}\, \frac{{\rm d}n}{{\rm d}L_{1.4}}\, \frac{L_{(1+z)\nu}(L_{1.4})}{4\pi}, \\
&S(M) = \exp\left(-\frac{M_{\rm cut}}{M}\right)\, \left[\int\, {\rm d}M\, \frac{{\rm d}n}{{\rm d}M}\, \exp\left(-\frac{M_{\rm cut}}{M}\right)\right]^{-1},
\end{align}
where $L_{1.4}$ is the radio luminosity at 1.4 GHz, ${\rm d}n/{\rm d}L_{1.4}$ is the luminosity function,
and $L_{(1+z)\nu}(L_{1.4})$ is given by Eq.~(\ref{eq:radio_luminosity}).

For the CIB, we adopt the model in \citet{2012MNRAS.421.2832S}:
\beq
\tilde{\cal J}_{\nu}(k|M) = f_{\nu, {\rm cen}}(M) + f_{\nu, {\rm sat}}(M) \tilde{u}_{\rm sat}(k|M),
\eeq
where $f_{\nu, {\rm cen}}(M)$ represents the luminosity-weighted HOD of central galaxies,
$f_{\nu, {\rm sat}}(M)$ is the HOD of satellites, and $\tilde{u}_{\rm sat}(k|M)$ is the Fourier transform of satellite number density profile.
The details of luminosity-weighted HOD are found in \citet{2012MNRAS.421.2832S}.



\bibliographystyle{mnras}
\bibliography{ref} 


\bsp	
\label{lastpage}
\end{document}